\documentclass[prd,twocolumn,floats,showpacs]{revtex4}
\usepackage{amssymb}
\usepackage{epsfig}
\usepackage{graphicx}
\usepackage{psfrag}

\def\D{\mathrm{d}} 
\def\E{\mathrm{e}}
\def\I{\mathrm{i}}

\begin{document}

\title{Neutrino dispersion in external magnetic fields}

\author{A.~V.~Kuznetsov$^{\rm (1)}$, 
N.~V.~Mikheev$^{\rm (1)}$,
G.~G.~Raffelt$^{\rm (2)}$,
L.~A.~Vassilevskaya$^{\rm (1)}$}
\affiliation{$^{\rm (1)}$Division of Theoretical Physics,
Department of Physics, Yaroslavl State University,\\
Sovietskaya 14, 150000 Yaroslavl, Russian Federation\\
$^{\rm (2)}$Max-Planck-Institut f\"ur Physik
  (Werner-Heisenberg-Institut), F\"ohringer Ring 6, 80805 M\"unchen,
  Germany} 

\date{15 October 2005}

\begin{abstract}
We calculate the neutrino self-energy operator $\Sigma (p)$ in the
presence of a magnetic field $B$.  In particular, we consider the
weak-field limit $e B \ll m_\ell^2$, where $m_\ell$ is the
charged-lepton mass corresponding to the neutrino flavor $\nu_\ell$,
and we consider a ``moderate field'' $m_\ell^2 \ll e B \ll m_W^2$.
Our results differ substantially from the previous literature.  For a
moderate field, we show that it is crucial to include the
contributions from all Landau levels of the intermediate charged
lepton, not just the ground state.  For the conditions of the early
universe where the background medium consists of a charge-symmetric
plasma, the pure $B$-field contribution to the neutrino dispersion
relation is proportional to $(e B)^2$ and thus comparable to the
contribution of the magnetized plasma.
\end{abstract}


\maketitle

\section{Introduction}

The presence of matter or electromagnetic fields modifies the
dispersion relation of neutrinos in rather subtle ways because these
elusive particles interact only by the weak force.  However,
Wolfenstein was the first to recognize that the feeble matter effect
is enough to affect neutrino flavor oscillations in dramatic ways
because the neutrino mass differences are very
small~\cite{Wolfenstein:1977}, with practical applications
in physics and astrophysics whenever neutrino oscillations
are important~\cite{Mohapatra:1998rq}.
The presence of external fields will lead to additional modifications
of the neutrino dispersion relation. There is a natural scale for the
field strength that is required to have a significant impact on
quantum processes, i.e.\ the critical value
\begin{equation}
B_e = m_e^2/e \approx 4.41 \times 10^{13}~\textrm{G}\,.
\end{equation}
Note that we use natural units where $\hbar=c=1$ and the
Lorentz-Heaviside convention where
$\alpha=e^2/4\pi\approx1/137$ so that $e\approx0.30>0$ is the
elementary charge, taken to be positive.

There are reasons to expect that fields of such or even larger
magnitudes can arise in cataclysmic astrophysical events such as
supernova explosions or coalescing neutron stars, situations where a
gigantic neutrino outflow should also be expected.  There are two
classes of stars, i.e.~soft gamma-ray repeaters
(SGR)~\cite{Kouveliotou:1999,Hurley:1999} and anomalous x-ray pulsars
(AXP)~\cite{Li:1999,Mereghetti:2002} that are believed to be remnants
of such cataclysms and to be magnetars~\cite{Duncan:1992}, neutron
stars with magnetic fields $10^{14}$--$10^{15}$~G.  The
possible existence of even larger fields of order
$10^{16}$--$10^{17}$~G is subject to
debate~\cite{Bisnovatyi-Kogan:1970, Bisnovatyi-Kogan:1989,
Balbus:1998, Akiyama:2003, Ardeljan:2004}.  The early universe between
the QCD phase transition (${}\sim 10^{-5}$~s) and the nucleosynthesis
epoch (${}\sim 10^{-2}$--$10^{+2}$~s) is believed to be yet another
natural environment where strong magnetic fields and large neutrino
densities could exist simultaneously~\cite{Grasso:2001}.

The modification of the neutrino dispersion relation in a magnetized
astrophysical plasma was studied in the previous
literature~\cite{D'Olivo:1989, Semikoz:1994, Elmfors:1996,
Erdas:1998}.  In particular, a charge-symmetric plasma with $m_e \ll T
\ll m_W$ and $B \lesssim T^2$ was considered for the early-universe
epoch between the QCD phase transition and big-bang nucleosynthesis.
Ignoring the neutrino mass, the dispersion relation for the electron
flavor was found to be~\cite{Elmfors:1996,Erdas:1998}
\begin{widetext}
\begin{equation}\label{eq:E_Raf}
\frac{E}{|{\bf p}|}=1+
\frac{\sqrt{2}\,G_{\rm F}}{3}
\left[
-\frac{7\pi^2T^4}{15} 
\left(\frac{1}{m_Z^2} + \frac{2}{m_W^2} \right)
+\frac{T^2eB}{m_W^2}\,\cos\phi
+\frac{(eB)^2}{2\pi^2m_W^2}\,
\ln\left(\frac{T^2}{m_e^2}\right)\,\sin^2 \phi\right],
\end{equation}
\end{widetext}
where $\bf p$ is the neutrino momentum and $\phi$ is the angle between
$\bf B$ and $\bf p$.  The first term in Eq.~(\ref{eq:E_Raf}) is the
dominating pure plasma contribution~\cite{Notzold:1988}, whereas the
second term is caused by the common influence of the plasma and
magnetic field~\cite{Elmfors:1996}.  The third term is of second order
in $(eB/T^2)\ll 1$ but was included because of the large logarithmic
factor $\ln(T/m_e)\gg 1$ \cite{Erdas:1998}.  The dispersion relation
of Eq.~({\ref{eq:E_Raf}}) applies to both $\nu_e$ and $\bar\nu_e$
without sign change in any of the terms \cite{footnote1}.


The $B$-field induced pure vacuum modification of the neutrino
dispersion relation was assumed to be negligible in these papers.
However, recently this contribution was calculated for the same
early-universe conditions as described
above~\cite{Elizalde:2002,Elizalde:2004}.  The dispersion relation
obtained in these papers for both $\nu_e$ and $\bar\nu_e$ can be
expressed as
\begin{equation}\label{eq:DE_Eliz}
\frac{E}{|{\bf p}|}=1+
\sqrt{2}\,G_{\rm F}\,
\frac{eB}{8\pi^2}\,\sin^2 \phi \, \E^{-p_\bot^2/(2eB)}\,,
\end{equation}
where $p_\bot$ is the momentum component perpendicular to the
$B$-field.  It is easy to check that this would be the dominant
$B$-field induced contribution by far and thus would lead to important
consequences for neutrino physics in
media~\cite{Ferrer:2000,Ferrer:2003,Ferrer:2004}.  The importance of
the question whether the $B$-field contribution to the neutrino
dispersion relation is dominant or negligible calls for an independent
calculation.

A literature search reveals that calculations of the neutrino
dispersion relation in external $B$-fields have a long
history~\cite{McKeon:1981,Borisov:1985,Erdas:1990}. To compare the
different results we introduce the neutrino self-energy operator
$\Sigma (p)$ that is defined in terms of the invariant amplitude for
the transition $\nu \to \nu$ by the relation
\begin{equation}\label{eq:sigma_def}
{\cal M}(\nu\to\nu)=-\bar\nu(p)\,\Sigma (p)\,\nu(p)\,,
\end{equation}
where $p$ is the neutrino four-momentum. Note that we use the
signature $(+,-,-,-)$ for the four-metric. 
Within the standard model, the general Lorentz
structure of $\Sigma(p)$ in the presence of a magnetic field can be
expressed in terms of four numerical coefficients $a$, $b$, $c$, and
$d$ as
\begin{equation}\label{eq:sigma}
\Sigma(p)=\left[a\,(p\gamma) + b\,(p \gamma)_{\|} + 
c \,\left(p \tilde\varphi \gamma \right) +
\I \,d \,\left(p \varphi \gamma \right) \right]\,L\,, 
\end{equation}
where $\gamma_\alpha$ are the Dirac matrices in the standard
representation and $L=\frac{1}{2}(1-\gamma_5)$ is the left-handed
projection operator.  The Lorentz indices of four-vectors and tensors
within parentheses are contracted consecutively.  For example,
$(p\varphi\gamma) = p^{\alpha}\varphi_{\alpha\beta}\gamma^{\beta}$.
Further, $\varphi$ is the dimensionless tensor of the 
electromagnetic field, normalized to the external $B$-field, whereas
$\tilde\varphi$ is its dual,
\begin{eqnarray}\label{eq:phi}
\varphi_{\alpha \beta} &=&  \frac{F_{\alpha \beta}}{B}\,,
\nonumber\\
{\tilde \varphi}_{\alpha \beta}&=&
\frac{1}{2} \varepsilon_{\alpha \beta \mu \nu} \varphi^{\mu \nu}\,. 
\end{eqnarray}
Finally, in the frame where only an external magnetic field $\bf B$ is
present, we take the spatial 3-axis to be directed along $\bf B$.
Four-vectors with the indices $\bot$ and $\|$ belong to the Euclidean
$\{1, 2\}$-subspace and the Minkowski $\{0, 3\}$-subspace,
correspondingly. For example, $p_\bot=(0,p_1,p_2,0)$ and
$p_\|=(p_0,0,0,p_3)$.  For any four-vectors $X$ and $Y$ we use the
notation
\begin{eqnarray}\label{eq:ABnotation}
(XY)_{\|} &=& (X \,\tilde \varphi \tilde \varphi \,Y) = 
X_0 Y_0 - X_3 Y_3\,,
\nonumber\\
(XY)_{\perp} &=& (X \,\varphi \varphi \,Y) =  X_1 Y_1 + X_2 Y_2\,, 
\nonumber\\
(XY) &=& (XY)_{\|} - (XY)_{\perp}\,.
\end{eqnarray}

\begin{table*}
\caption{\label{tab:coefficients}Coefficients in Eq.~(\ref{eq:sigma})
for the neutrino self-energy operator $\Sigma(p)$ in an external
$B$-field.}
\begin{ruledtabular}
\begin{tabular}{llllll}
Authors&Ref.&Field strength\footnote{$B$ is ``weak''
for $eB\ll m_\ell^2$ and ``moderate'' for $m_\ell^2\ll eB\ll m_W^2$.}
&$\displaystyle{b\times\frac{\sqrt{2}\,\pi^2}{G_{\rm F}}}$
&$\displaystyle{c\times\frac{\sqrt{2}\,\pi^2}{G_{\rm F}}}$&
$\displaystyle{d\times\frac{\sqrt{2}\,\pi^2}{G_{\rm F}}}$\\[2ex]
\hline
McKeon (1981)&\cite{McKeon:1981}&---&0&
$+3eB$&$+2eB$\\[4ex]
Erdas \& Feldman (1990)&\cite{Erdas:1990}&Moderate&
$\displaystyle{-\frac{(eB)^2}{3m_W^2}\,
\left(\ln\frac{m_W^2}{m_\ell^2}+\frac{3}{4}\right)}$
&0&0\\[4ex]
Elizalde et al.\ (2002)$^b$&\cite{Elizalde:2002}&Moderate&
$\displaystyle{+\frac{eB}{2}}$&
$\displaystyle{-\frac{eB}{2}}$&
0\\[4ex]
Elizalde et al.\ (2004)\footnote{Neutrino momentum range
$0 < p_{\perp}^2 \ll e B$.}
&\cite{Elizalde:2004}&Moderate&
$\displaystyle{+\frac{eB}{4}\,
\E^{-p_\bot^2/(2eB)}}$&
$\displaystyle{-\frac{eB}{4}\,
\E^{-p_\bot^2/(2eB)}}$&
0\\[4ex]
Our result (2005)\footnote{Neutrino momentum range
$0 < p_{\perp}^2 \ll m_W^2$.}&&Weak&
$\displaystyle{-\frac{(eB)^2}{3m_W^2}\,
\left(\ln\frac{m_W^2}{m_\ell^2}+\frac{3}{4}\right)}$&
$\displaystyle{+\frac{3eB}{4}}$&0\\[4ex]
Our result (2005)$^c$&&Moderate&
$\displaystyle{-\frac{(eB)^2}{3m_W^2}\,
\left(\ln\frac{m_W^2}{eB}+2.542\right)}$&
$\displaystyle{+\frac{3eB}{4}}$&
0\\[2ex]
\end{tabular}
\end{ruledtabular}
\end{table*}

\begin{figure}[htb]
\centering
\includegraphics{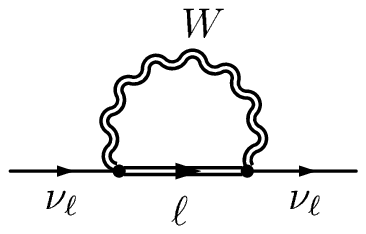}
\caption{Feynman diagram for the field-induced contribution to 
  the neutrino self-energy operator. Double lines denote exact
  propagators of the charged lepton and the $W$-boson in an external
  $B$ field. The contribution of the unphysical Higgs particles can be neglected
  in the limit $m_\ell^2 \ll m_W^2$.}
\label{fig:Feynman}
\end{figure}

Perturbatively, the matrix element of Eq.~(\ref{eq:sigma_def})
corresponds to the Feynman diagram shown in Fig.~\ref{fig:Feynman}
where double lines denote exact propagators in the external $B$ field.
Put another way, the self-energy operator corresponds to this Feynman
graph with the external neutrino lines truncated.
The motivation for our work is that the results obtained by different
authors at one-loop level do not agree with each other.  We anticipate
our results in Table~\ref{tab:coefficients} where we show $b$, $c$ and
$d$ obtained by previous authors and from our calculation detailed
below.

Turning to the interpretation of the coefficients in
Eq.~(\ref{eq:sigma}) we note that $a$ does not have an independent
meaning because for small neutrino energies, $E \ll m_W$, when $a (p)
= \textrm{const.}$, the first term in Eq.~(\ref{eq:sigma}) is fully
absorbed by the neutrino wave-function renormalization.  The
coefficient $d$ corresponds to an induced electric dipole moment and
as such can be non-zero only in the presence of the CP-odd field
invariant $(F\tilde F)=4\,{\bf E}\cdot {\bf B}$. Even in this case it
is strongly suppressed~\cite{Borisov:1985}.

Therefore, in our case of a pure external $B$-field only the
coefficients $b$ and $c$ are relevant for neutrino dispersion.  For a
massless neutrino the Dirac equation in momentum space is
\begin{equation}
\left[(p\gamma)-\Sigma(p)\right]\nu(p)=0\,.
\end{equation}
The dispersion relation follows from
\begin{equation}
\det\left[(p\gamma)-b(p \gamma)_{\|}L
-c(p\tilde\varphi\gamma)L\right]=0\,.
\end{equation}
For both $\nu$ and $\bar\nu$ this implies the same dispersion relation
\begin{equation}\label{eq:E(p)massless}
\frac{E}{|{\bf p}|} = 1 +
\left(b+\frac{c^2}{2}\right) \sin^2 \phi \,.
\end{equation}
Actually, in a perturbative sense the quantity $c^2$ is of higher
order, taking us to the two-loop level.  Therefore, to lowest order
the dispersion relation of massless neutrinos depends only on $b$.

For massive neutrinos the situation is more complicated because a
transverse $B$-field induces transitions between positive- and
negative-helicity states by the usual spin precession.  Put another
way, in the standard model neutrinos with mass inevitably have
magnetic dipole moments. In this situation it is not particularly
illuminating to express the effect of the $B$-field in terms of a
modified dispersion relation because an entirely new phenomenon
appears, the mixing of positive- with negative-helicity states.
Formally one can still proceed as above by $\Sigma(p)\to m+\Sigma(p)$
in the Dirac equation and obtain the new dispersion relation. However,
the energy eigenstates are no longer the left- and right-handed
helicity states but rather a superposition that depends on the
magnetic field orientation relative to ${\bf p}$.

Therefore, for massive neutrinos the effect of $B$ is better
illustrated in terms of the equation of motion of a free neutrino
state in an external homogeneous $B$-field. In this situation neutrino
and anti-neutrino states are not connected.  The Dirac equation
implies for the helicity amplitudes $\nu_\pm$ of a massive neutrino
\begin{equation}
{\rm i}\,\frac{\partial}{\partial t}
\left(\matrix{\nu_+\cr\nu_-\cr}\right)
=E_{\bf p}\left(1+\Omega_{\rm p}^B\right)
\left(\matrix{\nu_+\cr\nu_-\cr}\right)\,,
\end{equation}
where $E_{\bf p}=({\bf p}^2+m^2)^{1/2}$.  The dimensionless $B$-field
induced mixing matrix is
\begin{widetext}
\begin{eqnarray}\label{eq:Omega}
\Omega_{\rm p}^B&=&
\frac{1}{2}\,b\left[\sin^2\phi
\left(\matrix{1-v&0\cr0&1+v\cr}\right)
-\cos\phi\,\sin\phi\,\frac{v\,m}{E_{\bf p}}
\left(\matrix{0&1\cr1&0\cr}\right)
+\cos^2\phi\,\frac{m^2}{E_{\bf p}^2}
\left(\matrix{1&0\cr0&1\cr}\right)\right]\nonumber\\
&+&\frac{m\,c}{2E_{\bf p}}
\left[\sin\phi\left(\matrix{0&1\cr1&0\cr}\right)
-\cos\phi\,\frac{m}{E_{\bf p}}
\left(\matrix{1&0\cr0&-1\cr}\right)\right]\,,
\end{eqnarray}
\end{widetext}
where $v = |{\bf p}|/E$ is the neutrino velocity.  The first term
would represent at $v=1$ the energy shift for massless neutrinos of
Eq.~(\ref{eq:E(p)massless}) where we have dropped the higher-order
$c^2$ term. Of course, positive-helicity (right-handed) massless
neutrinos do not suffer an energy shift.

The second line is identical to the effect caused by a neutrino
magnetic moment
\begin{equation}
\mu_\nu = \frac{m\,c}{2 B}\,.
\end{equation}
Therefore, within the standard model the coefficient $c$ is implied by
the well-known result for the neutrino magnetic
moment~\cite{Lee:1977,Fujikawa:1980}
\begin{equation}
\mu_\nu = \frac{3e\,G_{\rm F}m}{8\pi^2\sqrt{2}}\,.
\end{equation}
Likewise, $c$ can be extracted from
Refs.~\cite{Borisov:1991a,Borisov:1991b}, where the neutrino
transitions $\nu_i \leftrightarrow \nu_j$ in an external
electromagnetic field were investigated.  From Eq.~(\ref{eq:Omega})
and Table~\ref{tab:coefficients} one concludes that for $m\gtrsim
10^{-4}\,\textrm{eV} \times (p_\bot/1\,\textrm{MeV})\,(B/B_e)$ the
neutrino energy shift becomes essentially non-diagonal.

If the neutrino mass is of Majorana type, the four neutrino components
in the full Dirac equation are not independent and the mass-implied
term vanishes---Majorana neutrinos do not have a magnetic moment.  On
the other hand, once neutrino masses are included one can not avoid
flavor mixing so that the situation becomes yet more complicated by
the presence of ordinatry flavor oscillations and flavor off-diagonal
magnetic spin precessions caused by magnetic transition moments which
exist for both Dirac and Majorana neutrinos.  We will not entangle our
discussion with these complications because in the standard model the
mass-induced effects all happen on the external neutrino legs in the
Feynman graph of Fig.~\ref{fig:Feynman}. The self-energy operator
$\Sigma(p)$ itself is independent of neutrino masses, at least to
one-loop order.

We begin in Secs.~\ref{sec:definition} and~\ref{sec:propagators} with
the technique to calculate the neutrino self-energy operator by using
the charged-lepton and $W$-boson propagators in a magnetic field.  In
Sec.~\ref{sec:Landaulevels} we calculate the neutrino self-energy
contribution from the $n$-th Landau level of the charged-lepton
propagator in combination with the exact $W$ propagator.  In contrast
to previous assumptions~\cite{Elizalde:2002,Elizalde:2004} we find
that it is not enough to use only the lowest Landau level.  In
Sec.~\ref{sec:sigma} we derive explicit results for the neutrino
self-energy operator in the limiting cases of a ``weak field'' $e B
\ll m_\ell^2$ and a ``moderate field'' $m_\ell^2 \ll e B \ll m_W^2$
before concluding in Sec.~\ref{sec:summary}.

\section{Definition of  \lowercase{\boldmath{$\Sigma(p)$}}}
\label{sec:definition}
                                                
The $\cal S$ matrix element for the transition $\nu \to \nu$
corresponds to the Feynman diagram shown in Fig.~\ref{fig:Feynman}
where double lines denote exact propagators in the presence of an
external magnetic field.  A detailed description of the calculational
techniques for quantum processes in external electromagnetic fields
can be found in Ref.~\cite{Kuznetsov:2003}.

For the charged lepton $\ell$ we consider a negative electric charge
$Q_\ell = - e < 0$. The propagator in the presence of a constant and
uniform magnetic field is translationally and gauge
non-invariant~\cite{Schwinger:1951}. It can be expressed as
\begin{equation}
S^{F}(x,y)  = \E^{\mbox{\normalsize $\I \Phi(x,y)$}}\,S (x-y)\,.
\label{eq:S0}
\end{equation}
Here, $S (x-y)$ is the translationally and gauge invariant part of the
propagator.  The translationally and gauge non-invariant phase can be
defined in terms of an integral along an arbitrary contour as
\begin{equation}
\Phi (x,y) =  - e \,
\int_x^y
\D \xi_\mu \, K^\mu (\xi)\,,
\label{eq:FKA}
\end{equation}
where $K^\mu (\xi)=A^\mu(\xi)+\frac{1}{2}F^{\mu \nu}(\xi - y)_\nu$.

The corresponding $W$-boson propagator can be represented in a
similar form
\begin{equation}
G^F_{\rho \sigma} (x,y)  = \E^{\mbox{\normalsize $\I \Phi (x,y)$}} \,
G_{\rho \sigma} (x-y)\,,
\label{eq:G0}
\end{equation}
where $\Phi (x,y)$ is also given by Eq.~(\ref{eq:FKA}).

It is useful to consider the Fourier transforms of the translationally
invariant parts of the propagators
\begin{eqnarray}
S (X) &=&
\int \frac{\D^4 q}{(2 \pi)^4} \, S (q) \, \E^{- \I q X}\,,
\label{eq:S_Four}\\
&&\nonumber\\
G_{\rho \sigma} (X) &=&
\int \frac{\D^4 q}{(2 \pi)^4} \, G_{\rho\sigma} (q)\,\E^{- \I q X}\,.
\label{eq:G_Four}
\end{eqnarray}

The sum of the translationally non-invariant phases of the lepton and
$W$ propagators Eqs.~(\ref{eq:S0}) and~(\ref{eq:G0}) in the loop 
vanishes,
\begin{equation}
\Phi (x,y)  + \Phi (y,x) = 0\,.
\end{equation}
This allows one to extract the amplitude $\cal M$ from the
$\cal S$ matrix element by the standard method,
\begin{eqnarray}\label{eq:M_int}
{\cal M} (\nu \to \nu) &=& \frac{\I\,g^2}{2}\, 
\bar\nu(p) \, \gamma^\alpha L 
\int\frac{\D^4 q}{(2 \pi)^4} \,S (q)\,G_{\beta \alpha} (q-p)
\nonumber\\
&&\kern8em{}\times
\gamma^\beta L\,\nu (p)\,,
\end{eqnarray}
where $g$ is the electroweak $SU (2)$ coupling constant of the
standard model.  Comparing this result with Eq.~(\ref{eq:sigma_def})
one can express the neutrino self-energy operator as
\begin{equation}
\Sigma (p) = - \frac{\I\,g^2}{2} \, \gamma^\alpha \, 
L \, J_{\alpha \beta} (p) \, \gamma^\beta \, L \,,
\label{eq:sigma1}
\end{equation}
where
\begin{equation}
J_{\alpha \beta} (p) =
\int \, \frac{\D^4 q}{(2 \pi)^4} \,S (q) \, G_{\beta \alpha} (q-p) \,.
\label{eq:J_alpha_beta_def}
\end{equation}
To calculate $J_{\alpha \beta} (p)$ we need to unravel the
propagators.

\hbox{\ }

\section{Charged-lepton and \boldmath{$W$}-boson propagator in a
magnetic field}                                \label{sec:propagators}

For the Fourier transform $S (q)$ of the translationally invariant
part of the lepton propagator Eq.~(\ref{eq:S_Four}) one obtains in the
Fock proper-time formalism~\cite{Schwinger:1951}
\begin{widetext}
\begin{equation}\label{eq:S(q)}
S (q) = \int_0^{\infty}\!\! 
\frac{\D s}{\cos (\beta s)}\,
\exp\left[- \I s \left(m_\ell^2- q_{\|}^2 
+ \frac{\tan (\beta s)}{\beta s}\,q_{\perp}^2 \right) \right] 
\biggl\{\left[(q \gamma)_{\|} + m_\ell\right] \left[
\cos (\beta s) - \frac{ (\gamma \varphi \gamma)}{2} \,
\sin (\beta s) \right]-
\frac{(q \gamma)_{\perp}}{\cos (\beta s)}\biggr\},
\end{equation}
where $\beta = e B$ and $m_\ell$ is the lepton mass.  Similarly, the
Fourier transform of the translationally invariant part of the
$W$-boson propagator Eq.~(\ref{eq:G_Four}) can be written in Feynman
gauge as~\cite{Erdas:1990}

\begin{equation}\label{eq:G(q)}
G_{\rho \sigma} (q) = - \int_0^{\infty} 
\frac{\D s}{\cos (\beta s)} 
\exp\left[- \I s \left(m_W^2- q_{\|}^2 
+ \frac{\tan (\beta s)}{\beta s}\,q_{\perp}^2 \right) \right] 
\biggl[
(\tilde\varphi \tilde\varphi)_{\rho \sigma} 
- (\varphi \varphi)_{\rho \sigma} \, \cos (2 \beta s)
- \varphi_{\rho \sigma} \, \sin (2 \beta s) \biggr] .
\end{equation}
Manipulations with the exact expressions Eqs.~(\ref{eq:S(q)}) and
(\ref{eq:G(q)}) are extremely cumbersome.  Magnetic fields existing in
Nature probably are always weak compared with the critical field for
the $W$-boson, $m_W^2/e \simeq 10^{24}\,\textrm{G}$.  Therefore, the
$W$ propagator can be expanded in powers of $\beta$ as a small
parameter. We find up to second order
\begin{equation} 
G_{\rho \sigma} (q)=
- \I \, \frac{g_{\rho \sigma}}{q^2 - m_W^2} \, - 
\, \beta \, \frac{2 \, \varphi_{\rho \sigma}}{(q^2 - m_W^2)^2} \, 
+ 
\I\, \beta^2 \biggl[g_{\rho \sigma} \left(\frac{1}{(q^2 - m_W^2)^3} + 
\frac{2 \, q_{\perp}^2}{(q^2 - m_W^2)^4} \right) 
+ 
4 \, (\varphi \varphi)_{\rho \sigma} \,\frac{1}{(q^2 - m_W^2)^3} 
\biggr]
+ {\cal O}(\beta^3)\,.
\label{eq:G(q)<}
\end{equation}

Likewise, the asymptotic expression for the lepton propagator $S (q)$
is realised when the field strength is the smallest dimensional parameter, 
$\beta \ll m_\ell^2 \ll m_W^2$.  In this ``weak field approximation'' the
charged-lepton propagator can be expanded as~\cite{Chyi:2000}
\begin{equation}\label{eq:S(q)<} 
S (q)=
\I \, \frac{(q \gamma) + m_\ell}{q^2 - m_\ell^2} \, + 
\, \beta \, \frac{(q \gamma)_{\|} + m_\ell}{2 (q^2 - m_\ell^2)^2} 
\,(\gamma \varphi \gamma)
+\beta^2 \, \frac{2 \, \I
\left[(q_{\|}^2 - m_\ell^2) \, (q \gamma)_{\perp} 
- q_{\perp}^2 \, ((q \gamma)_{\|} + m_\ell) \right]}
{(q^2 - m_\ell^2)^4} 
+ {\cal O}(\beta^3)\,.
\end{equation}
One can see from this expansion that the contribution of the region of
small virtual momenta $q^2 \sim m_\ell^2 \ll m_W^2$ is enhanced in
each succeeding term. If the propagator is used for a ``moderate
field,'' $m_\ell^2 \ll \beta \ll m_W^2$, the expansion is not
applicable and the exact propagator Eq.~(\ref{eq:S(q)}) must be used.

When the magnetic field is strong enough, $B \gtrsim B_\ell =
m_\ell^2/e$, another possibility is to express the charged-lepton
propagator as an expansion in terms of Landau
levels~\cite{Chodos:1990}
\begin{equation} 
S (q) = \sum\limits_{n=0}^\infty \, 
\frac{\I}{q_{\|}^2 - m_\ell^2 - 2 n \beta}
\biggl\{\left[(q \gamma)_{\|} + m_\ell\right] 
\left[d_n (v) - \frac{\I}{2} \, 
(\gamma \varphi \gamma) \, d_n^\prime(v)\right] 
-(q \gamma)_{\perp} \, 2 n \, \frac{d_n (v)}{v} \biggr\} ,
\label{eq:S(q)>}
\end{equation}
\end{widetext}
where $v = q_{\perp}^2/\beta$ and 
\begin{equation}
d_n (v) = (-1)^n \E^{-v} [L_n (2 v) - L_{n-1} (2 v)]\,.
\label{eq:d_fun}
\end{equation}
Here, $L_n (x)$ are the Laguerre polynomials with the additional
definition $L_{-1} (x) = 0$.

\section{Contribution of the lepton low Landau levels}
\label{sec:Landaulevels}

As we have already stressed in the introduction, our final result
derived in Sec~\ref{sec:sigma} below
strongly disagrees with that of
Refs.~\cite{Elizalde:2002,Elizalde:2004}.  We think that the
disagreement arises because these authors use only one lowest Landau
level in the charged-lepton propagator in the case of moderate field
strengths which they call ``strong fields.''  However, the
contributions of the next Landau levels can be of the same order as
the ground-level contribution because in the integration over the
virtual lepton four-momentum in the loop the region $q_{\|}^2 \sim
m_W^2 \gg \beta$ appears to be essential.  

To substantiate this point we calculate the contribution to the
neutrino self-energy operator from the $n$-th charged-lepton Landau
level in conjunction with the exact $W$-propagator in the limit
$p_{\perp}^2/m_W^2 \ll m_W^2/\beta$.
Substituting the exact $W$-propagator Eq.~(\ref{eq:G(q)}) and
the $n$-th charged-lepton Landau level from Eq.~(\ref{eq:S(q)>}) into
Eq.~(\ref{eq:J_alpha_beta_def}) we find
\begin{widetext}
\begin{eqnarray}\label{eq:J^n1}
J_{\sigma \rho}^{(n)}(p)&=& - \int\frac{\D^4 q}{(2 \pi)^4}\,
\frac{\I}{q_{\|}^2 - m_\ell^2 - 2 n \beta}
\biggl\{(q \gamma)_{\|} 
\left[d_n (v) - \frac{\I}{2} \, (\gamma \varphi \gamma) \
\,d_n^\prime(v)\right] 
-(q \gamma)_{\perp} \, 2 n \, \frac{d_n (v)}{v} \biggr\}
\nonumber\\
&\times&
\int_0^{\infty}\!\!
\frac{\D s}{\cos (\beta s)} \,
\exp\biggl[- \I s \left(m_W^2- (q-p)_{\|}^2 
+ \frac{\tan (\beta s)}{\beta s}\,(q-p)_{\perp}^2 \right) \biggr]  
\biggl[
(\tilde\varphi \tilde\varphi)_{\rho \sigma} 
- (\varphi \varphi)_{\rho \sigma} \, \cos (2 \beta s)
- \varphi_{\rho \sigma} \, \sin (2 \beta s) \biggr].
\nonumber\\
\end{eqnarray}
Terms with even numbers of $\gamma$ matrices were omitted because they
are removed by the chiral structure of the operator
Eq.~~(\ref{eq:sigma1}).  Next we perform a clockwise rotation in the
complex plane $s = - \I \tau$ and use the identity
\begin{equation}
\frac{1}{q_{\|}^2 - m_\ell^2 - 2 n \beta} 
= - \int_0^{\infty}
\D \tau' \, \exp \left[- \tau' \left( m_\ell^2 + 2 n \beta 
- q_{\|}^2 \right) \right]\,.
\label{eq:int_tau'}
\end{equation}
These manipulations allow us to rewrite the integral
Eq.~(\ref{eq:J^n1}) as
\begin{eqnarray}\label{eq:J^n2} 
J_{\sigma \rho}^{(n)} (p)&=&\int\frac{\D^4 q}{(2 \pi)^4} \,
\biggl\{(q \gamma)_{\|} 
\left[d_n (v) - \frac{\I}{2} \, 
(\gamma \varphi \gamma) \, d_n^\prime(v)\right] 
- (q \gamma)_{\perp} \, 2 n \, \frac{d_n (v)}{v} \biggr\}
\nonumber\\
&\times&
\int_0^{\infty} 
\frac{\D \tau \, \D \tau'}{\cosh (\beta \tau)}\, 
\biggl[(\tilde\varphi \tilde\varphi)_{\rho \sigma} 
- (\varphi \varphi)_{\rho \sigma} \, \cosh (2 \beta \tau)
+ \I \, \varphi_{\rho \sigma} \, \sinh (2 \beta \tau) \biggr] 
\nonumber\\
&\times& 
\exp\biggl[- \tau' \left(m_\ell^2 + 2 n \beta - q_{\|}^2\right)
- \tau \left(m_W^2- (q-p)_{\|}^2 \right) 
- \frac{\tanh (\beta \tau)}{\beta}\,(q-p)_{\perp}^2 \biggr]\,.
\end{eqnarray}
In the integration over $\D^4 q = \D^2 q_{\|} \, \D^2 q_{\perp}$, the
integrals over $\D^2 q_{\|}$ can be easily calculated because they are
of Gaussian form.  As a result we find
\begin{eqnarray} 
J_{\sigma \rho}^{(n)} (p)&=&
\frac{\I}{16 \pi^3 m_W^2}\int_0^{\infty} 
\frac{\D x \, \D y}{(x+y) \cosh (\eta x)}\,
\exp\left[- x  + \xi \frac{xy}{x+y} - y \,(2 n \eta + \lambda) \right]
\nonumber\\
&\times& 
\biggl[(\tilde\varphi \tilde\varphi)_{\rho \sigma} 
- (\varphi \varphi)_{\rho \sigma} \, \cosh (2 \eta x)
+ \I \, \varphi_{\rho \sigma} \, \sinh (2 \eta x) \biggr]
\nonumber\\
&\times& 
\int \, \D^2 q_{\perp} \,
\exp\left[- \frac{\tanh (\eta x)}{\beta}\,(q-p)_{\perp}^2 \right]
\biggl\{(p \gamma)_{\|} \, \frac{x}{x+y}
\left[d_n (v) - \frac{\I}{2} \, (\gamma \varphi \gamma)\,d_n^\prime
  (v)\right]
-(q \gamma)_{\perp} \, 2 n \, \frac{d_n (v)}{v} \biggr\} ,
\label{eq:J^n3}
\end{eqnarray}
%
where the dimensionless variables $x = m_W^2 \tau$ and $y = m_W^2
\tau'$ have been introduced as well as the parameters $\eta =
\beta/m_W^2$, $\xi = p_{\|}^2/m_W^2 \simeq p_{\perp}^2/m_W^2$ and
$\lambda = m_{\ell}^2/m_W^2$.  From Eq.~(\ref{eq:J^n3}) follows that
the essential region of the $x$ variable is $x \sim 1$ due to the
exponential $\E^{-x}$. Given the condition $\eta \ll 1$, the argument
of the hyperbolic functions is small, $\eta x \ll 1$, leading to an
obvious simplification. One should also take into account the
condition $q_{\perp}^2 \sim \beta$ caused by the 
functions $d_n (v)$, see Eq.~(\ref{eq:d_fun}), containing the 
exponential $\E^{-v}$. For a
wide range of the numbers $n$ the exponential in the integral over
$\D^2 q_{\perp}$ is simplified, with the only restriction $n \ll
1/\eta = m_W^2/\beta$:
\begin{eqnarray} 
\exp\left[- \frac{\tanh (\eta x)}{\beta}\,(q-p)_{\perp}^2 \right] 
\simeq \exp\left(- x\,\frac{p_{\perp}^2}{m_W^2}\right) 
\,\times \,
\exp\left(- x\,\frac{q_{\perp}^2 - 2(qp)_{\perp}}{m_W^2}\right) .
\label{eq:simpl_exp}
\end{eqnarray}
Here, the first exponential is equal to $\E^{- \xi x}$. We consider
the value $p_{\perp}^2$ to vary in a very wide range, $0 < p_{\perp}^2
\ll m_W^4/\beta$.  The second exponential is equal to unity with a
good accuracy, because $q_{\perp}^2 \sim \beta \ll m_W^2$ and
$(qp)_{\perp} \ll m_W^2$.  With these approximations, the integration
over $\D^2 q_{\perp}$ can be easily performed,
\begin{eqnarray} 
\int \, \D^2 q_{\perp} \,
d_n (v) = \pi \, \beta \, (2 - \delta_{n0})\,,\quad
%
\int \, \D^2 q_{\perp} \,
d_n\,' (v) = - \pi \, \beta \, \delta_{n0}\,, \quad
%
\int \, \D^2 q_{\perp} \,
(q \gamma)_{\perp} \, \frac{d_n (v)}{v} = 0\,.
\label{eq:int_q_perp}
\end{eqnarray}
The investigated integral acquires the form
%
\begin{equation}\label{eq:J^n4}
J_{\sigma \rho}^{(n)} (p)= 
\frac{\I\,\beta}{16 \pi^2 m_W^2} \, (p \gamma)_{\|} \, g_{\rho \sigma}
\left\{ 2 - \left[1 - \frac{\I}{2} \,
(\gamma \varphi \gamma)\right] \delta_{n0}\right\}
\int_0^{\infty} 
\frac{x \,\D x \, \D y}{(x+y)^2}\,
\exp\left[- x  - \xi \frac{x^2}{x+y} - y \,
(2 n \eta + \lambda) \right].
\end{equation}
Taking into account the smallness of the parameters $\eta$ and
$\lambda$, one finally obtains for $n \ll m_W^2/\beta$
\begin{equation}\label{eq:J^n5}
J_{\sigma \rho}^{(n)} (p) = 
\frac{\I\,\beta}{16 \pi^2 \, p_{\perp}^2}
\, \ln \left( 1 + \frac{p_{\perp}^2}{m_W^2} \right)
(p \gamma)_{\|} \, g_{\rho \sigma}
\left\{ 2 - \left[1 - \frac{\I}{2} \,
(\gamma \varphi \gamma)\right] \delta_{n0}
\right\}.
\end{equation}
Substituting Eq.~(\ref{eq:J^n5}) into Eq.~(\ref{eq:sigma1}) we finally
find the contribution of the $n$-th Landau level of the lepton
propagator to the neutrino self-energy operator
\begin{equation}\label{eq:M_n_W}
\Sigma^{(n)} (p)=- \frac{G_{\rm F} \, e B}{\sqrt{2}\;2 \pi^2}
\,\frac{m_W^2}{p_{\perp}^2}
\,\ln \left( 1 + \frac{p_{\perp}^2}{m_W^2} \right)
\left[ (2 - \delta_{n0}) \, (p \gamma)_{\|} - \delta_{n0} \,
(p \tilde\varphi \gamma)\right] \, L \,.
\end{equation}
We conclude from Eq.~(\ref{eq:M_n_W}) that, contrary to the treatment
of Refs.~\cite{Elizalde:2002,Elizalde:2004}, the lowest Landau level
does not dominate.

For higher Landau levels, $n \gtrsim m_W^2/\beta$, the calculation is
more cumbersome.  Therefore, using the lepton propagator expanded
in terms of the Landau levels, with a further summation, is extremely
inconvenient.  It is much simpler to take the exact lepton propagator
in the form of Eq.~(\ref{eq:S(q)}). This approach is used in 
Sec~\ref{sec:sigma} below. 

In Ref.~\cite{Elizalde:2004} a numerical test of the lowest Landau
level domination was made. However, we believe that the results of
this test are misleading.  In the right-hand side of Fig.~2 of
Ref.~\cite{Elizalde:2004}, corresponding to strong fields $e B \simeq
10^4 m_e^2$, the value $\varkappa_a$ defined in Eq.~(89) of
Ref.~\cite{Elizalde:2004} tends to unity.  From this behavior it was
concluded that the lowest Landau level domination worked well in this
region. However, the left-hand side of that plot, corresponding to
weak fields $e B \simeq 10^{-4} m_e^2$, would have to coincide with,
but strongly contradicts, the weak-field result of
Ref.~\cite{Erdas:1990}.  An accurate analysis of Eqs.~(89) and~(90) of
Ref.~\cite{Elizalde:2004} shows that a precise cancellation of the two
infinities arises, whereas the rest appears to be of order $e B/m_W^2
\lesssim 10^{-6}$ for the relevant field values, but not of order
unity as claimed there.

\section{New calculation of \lowercase{\boldmath{$\Sigma(p)$}}}
\label{sec:sigma}
                                                     
We begin our calculation of $\Sigma (p)$ with the simpler case of a
``weak field'' where $B$ defines the smallest energy scale of the
problem, $e B \ll m_\ell^2 \ll m_W^2$.  As an essential simplification
we use the field expansions of the Fourier transformed $W$ and lepton
propagators.  Substituting Eqs.~(\ref{eq:S(q)<}) and~(\ref{eq:G(q)<})
in Eq.~(\ref{eq:J_alpha_beta_def}), and assuming in addition that 
$p_{\perp}^2 \ll m_W^2$, 
we find an expansion of $J_{\alpha \beta} (p)$ in powers of the field strength,
%
\begin{eqnarray}\label{eq:J<_res}
J_{\alpha \beta} (p)&=& 
\frac{1}{16 \pi^2}
\biggl\{ \frac{e B}{m_W^2} \left[ - \frac{\I}{2} \,
g_{\alpha \beta} \,
(p \tilde\varphi \gamma) \gamma_5 + \varphi_{\alpha \beta} \,
(p \gamma) \right]\nonumber\\
&&\kern2.5em{}+
 \I \left(\frac{e B}{m_W^2}\right)^2 \biggl[ g_{\alpha \beta} \,
(p \gamma)_{\|} \left(\frac{2}{3} \ln \frac{m_W^2}{m_\ell^2}
- \frac{3}{2} \right)
+ \I \, \varphi_{\alpha \beta} \, (p \tilde\varphi \gamma) \gamma_5
- (\varphi \varphi)_{\alpha \beta} \,(p \gamma) \biggr]\biggr\}
 + \ldots\,.
\end{eqnarray}
The dots include terms having the structure $g_{\alpha \beta} \,(p
\gamma)$ and contain, in particular, the ultraviolet divergence, to be
fully absorbed by the neutrino wave-function renormalization.  They
also include terms with an even number of $\gamma$ matrices that are
removed by the chiral structure of $\Sigma(p).$

Substituting Eq.~(\ref{eq:J<_res}) into Eq.~(\ref{eq:sigma1}) we
finally find
\begin{equation}
\Sigma (p) = \frac{G_{\rm F}}{\sqrt{2}\,4 \pi^2}
\biggl[3 e (p \tilde F \gamma)
-\frac{e^2 (p \tilde F \tilde F \gamma)}{m_W^2} 
\left(\frac{4}{3} \ln \frac{m_W^2}{m_\ell^2} + 1 \right)
\biggr] \, L \,.
\label{eq:M_mu}
\end{equation}
The corresponding coefficients $b$ and $c$ are shown in
Table~\ref{tab:coefficients}.

For a ``moderate field,'' $m_\ell^2 \ll e B \ll m_W^2$ we use the
exact expression Eq.~(\ref{eq:S(q)}) for the charged-lepton propagator
and the expansion Eq.~(\ref{eq:G(q)<}) for the $W$ propagator, with the 
same assumption $p_{\perp}^2 \ll m_W^2$.  After
a straightforward but cumbersome calculation we find
\begin{eqnarray}\label{eq:J>_res} 
J_{\alpha \beta} (p) &=& 
\frac{1}{16 \pi^2} \biggl\{ \frac{e B}{m_W^2}
\left[ - \frac{\I}{2} \, g_{\alpha \beta} \,
(p \tilde\varphi \gamma) \gamma_5 + \varphi_{\alpha \beta}
\, (p \gamma) \right] 
\nonumber\\
&+& \I \left(\frac{e B}{m_W^2}\right)^2 \biggl[ g_{\alpha \beta} \,
(p \gamma)_{\|} \biggl(\frac{2}{3} \ln \frac{m_W^2}{e B} 
- \frac{7}{6} 
+\frac{1}{3} \, \ln 3 + \frac{2}{3} \, \gamma_{\rm E} - 2 I \biggr) 
+\I \, \varphi_{\alpha \beta} \, (p \tilde\varphi \gamma) \gamma_5
- (\varphi \varphi)_{\alpha \beta} \, (p \gamma) \biggr] \biggr\}
 + \dots \,,\nonumber\\
\end{eqnarray}
%
where $\gamma_{\rm E} = 0.577\dots$ is the Euler constant, and
\begin{equation}
I = \int_0^\infty \, \frac{\D z}{z^3}
\left(\frac{z^2}{\sinh^2 z} - \frac{3}{3 + z^2} \right)
\approx -0.055.
\label{eq:int_C}
\end{equation}
The presence of the term $\beta^2 \ln \beta$ with $\beta = e B$ in
Eq.~(\ref{eq:J>_res}) shows that an expansion of the lepton propagator
in powers of $\beta$ as a small parameter is not possible.

Finally we find 
%
\begin{equation}
\Sigma (p) = \frac{G_{\rm F}}{\sqrt{2}\,4 \pi^2}
\biggl[ 3 e (p \tilde F \gamma)
- \frac{e^2 (p \tilde F \tilde F \gamma)}{m_W^2} \left(\frac{4}{3} 
\ln \frac{m_W^2}{e B} + 3.389 \right)
\biggr] \, L \,.
\label{eq:M_e}
\end{equation}
Again, the corresponding coefficients $b$ and $c$ are shown in
Table~\ref{tab:coefficients}.
\eject
\end{widetext}

\section{Summary}                                  \label{sec:summary}

We have calculated the neutrino self-energy operator $\Sigma (p)$ in a
magnetic field at one-loop order.  Our results for the invariant
coefficients of Eq.~(\ref{eq:sigma}) that characterize $\Sigma (p)$
are shown in Table~\ref{tab:coefficients} for weak and moderate field
strengths and are compared to those of previous authors. Our results
strongly disagree with those of the recent
Refs.~\cite{Elizalde:2002,Elizalde:2004} where a large effect was
found.  For moderate fields we have shown that considering only the
lowest Landau level contribution in the lepton propagator is incorrect
because the contributions of the next Landau levels are of similar
magnitude.

It is instructive to reproduce explicitly the energy of a $\nu_e$ or
$\bar\nu_e$ in the presence of a CP symmetric plasma and the
simultaneous presence of a magnetic field.  We write the energy shift
in the form of Eq.~(\ref{eq:E_Raf}).  For a weak field $e B \ll m_e^2$
we find
\begin{widetext}
\begin{equation}
\frac{E}{|{\bf p}|}=1+
\frac{\sqrt{2}\,G_{\rm F}}{3}\,
\left[-\frac{7\pi^2T^4}{15}
\left(\frac{1}{m_Z^2} + \frac{2}{m_W^2} \right)
+\frac{T^2eB}{m_W^2}\,\cos\phi 
+\frac{(e B)^2}{2\pi^2m_W^2}
\,\sin^2 \phi 
\left( \ln \frac{T^2}{m_e^2} - \ln \frac{m_W^2}{m_e^2} - 
\frac{3}{4}\right)
\right]\,.
\end{equation}
\end{widetext}
Interestingly, the logarithmic $B$-field induced plasma term in the
third term and the logarithmic pure $B$-field term add to
$\ln(T^2/m_W^2)$ so that no electron-mass dependence remains.  For a
moderate field $m_e^2 \ll e B \ll m_W^2$ the third term is
\begin{equation}
+\frac{(e B)^2}{2\pi^2m_W^2}
\,\sin^2 \phi 
\left( \ln \frac{T^2}{m_e^2} - \ln \frac{m_W^2}{eB}-2.542 \right)\,.
\end{equation}
In this case an electron-mass dependence remains.

In a plasma, the pure $B$-field term is comparable to the logarithmic
contribution of the $B$-field induced plasma term derived in
Ref.~\cite{Erdas:1998}.  However, these logarithmic terms do not seem
to be numerically important relative to the term linear in $eB$.

\section*{ACKNOWLEDGMENTS}

We thank V.~A.~Rubakov for useful discussions.  The work of
A.~V.~Kuznetsov and N.~V.~Mikheev was supported in part by the Russian
Foundation for Basic Research under Grant No.~04-02-16253 and by the
Council on Grants by the President of the Russian Federation for the
Support of Young Russian Scientists and Leading Scientific Schools of
the Russian Federation under Grant No.~NSh-1916.2003.2.  G.~G.~Raffelt
acknowledges partial support by the Deutsche Forschungsgemeinschaft
under Grant No.~SFB-375 and by the European Union under the
Ilias project, contract No.~RII3-CT-2004-506222.


\end{document}